\def\rfr#1{eq. (\ref{#1})}

\def\bm#1{{\mbox{\boldmath$#1$\unboldmath}}}
\def\asec{$''$ cy$^{-1}$}

\def\dert#1#2{\frac{{{d}}{#1}}{{{d}}{#2}}}              

\def\asec{$''$ cy$^{-1}$}

\def\bar{\begin{eqnarray}}
\def\ear{\end{eqnarray}}
\def\bb{\bibitem}
\def\eqI{\begin{equation}}
\def\eqF{\end{equation}}
\def\eqIa{\begin{eqnarray}}
\def\eqFa{\end{eqnarray}}
\def\rp#1#2{{#1\over#2}}

\def\lb#1{\label{#1}}
\def\beq{\begin{equation}}
\def\eeq{\end{equation}}

\def\oc2{$\mathcal{O}(c^{-2})$}

\def\bds#1{\boldsymbol{#1}}

\documentclass{aastex}
\usepackage{amsmath,amsthm,amscd,amssymb}
\usepackage{latexsym}
\usepackage{graphicx,epsfig}

\RequirePackage{color}

\begin{document}

\title{Gravitomagnetic effects in Kerr-de Sitter space-time}

\shorttitle{Gravitomagnetic effects in Palatini $f(R)$ gravity}
\shortauthors{L. Iorio and M.L. Ruggiero}

\author{Lorenzo Iorio }
\affil{INFN-Sezione di Pisa. Permanent address for correspondence: Viale Unit\`{a} di Italia 68, 70125, Bari (BA), Italy. E-mail: lorenzo.iorio@libero.it}
\and
\author{Matteo Luca Ruggiero}
\affil{Dipartimento di Fisica del Politecnico di Torino and
INFN-Sezione di Torino, Corso Duca degli Abruzzi 24, 10129, Torino
(TO), Italy. E-mail: matteo.ruggiero@polito.it}

\begin{abstract}
We explicitly worked out the orbital effects induced on the
trajectory of a test particle by the the weak-field approximation
of the Kerr-de Sitter metric. It results that the
node, the pericentre and the mean anomaly undergo secular
precessions proportional to $k$, which is a measure of the non
linearity of the theory. We used such theoretical predictions and
the latest observational determinations of the non-standard
precessions of the perihelia of the inner planets of the Solar
System to put a bound on $k$ getting $k\leq 10^{-29}$ m$^{-2}$. The node rate of the LAGEOS Earth's satellite yields
$k\leq 10^{-26}$ m$^{-2}$. The periastron precession of the double pulsar PSR J0737-3039A/B allows to obtain $k\leq 3\times 10^{-21}$ m$^{-2}$. Interpreting $k$ as a cosmological constant $\Lambda$, it turns out that such constraints are weaker than those obtained from the Schwarzschild-de Sitter metric.

\end{abstract}

\section{Introduction} \label{sec:intro}

The General Theory of Relativity (GTR) has passed with excellent results many
observational tests, as Solar System and binary pulsars
observations show \citep{Wei,Will06,Tur08}.  As a matter of fact, the current values of the PPN parameters are in
agreement with GTR predictions.

However, some observations seem to
question the general relativistic model of gravitational
interaction on larger scales. On the one hand, the data coming
from the galactic rotation curves of spiral galaxies
\citep{Binney87} cannot be explained on the basis of Newtonian
gravity or GTR: the existence of \textit{dark matter} is postulated
to reconcile the theoretical model with observations; furthermore,
dark matter can explain the mass discrepancy in galactic clusters
\citep{Clowe06}. On the other hand, a lot of observations, such as
the light curves of the type Ia supernov{\ae} and the cosmic
microwave background (CMB) experiments
\citep{Riess98,Perlmutter99,Bennet03}, firmly state that our
Universe is now undergoing a phase of accelerated expansion.
Actually, the present acceleration of the Universe cannot be
explained, within GTR, unless the existence of a cosmic fluid
having exotic properties is postulate, i.e. the so called
\textit{dark energy}.

The cosmological constant is one of the candidates to explain (in the GTR framework) what the dark energy is (see e.g. \citet{peebles03} and references therein).
On the other hand, modified gravity models that go beyond GTR have been proposed to try to explain current observations and, among these models, $f(R)$ theories of
gravity \citep{sotfar08} received much attention in recent years. In these theories
the gravitational lagrangian depends on an arbitrary function $f$
of the scalar curvature $R$;  they are also referred to as
``extended theories of gravity'', since they naturally generalize,
on a geometric ground, GTR: namely, when
$f(R)=R$ the action reduces to the usual Einstein-Hilbert action,
and Einstein's theory is obtained. It is interesting to point out that
the  vacuum solutions of GTR with a cosmological
constant are also solutions of $f(R)$ gravity vacuum field equations: this is always true in the Palatini formalism,
while in metric $f(R)$ gravity this holds for the solutions \textit{with constant scalar curvature
$R$} \citep{FFVa,allemandi05,magnano}.

The relevance of the cosmological constant   in modern gravitational physics is manifest,  and it  is interesting  to focus on the solutions of Einstein's field equations with cosmological constant, to investigate its role on different scales. For instance, the Schwarzschild-de Sitter metric, which describes a point-like mass in a space-time with a cosmological constant,  has
been recently studied by \citet{kagramanova06,Sereno,Jetzer,IorioL}. In particular, the Schwarzschild-de Sitter metric has been studied to investigate the influence of the cosmological constant on gravitational lensing in \citet{rindler07,sereno08a,ruggiero07,sereno08b}.

In this paper we are concerned with the Kerr-de Sitter metric, which describes
a rotating black-hole in a space-time with a cosmological constant
\citep{demianski73,carter73,kerr03,kra04,kra05,kra07}. In particular, we want to study the gravito-magnetic (GM) effects in
Kerr-de Sitter metric. GM effects are due to the rotation of
the sources of the gravitational field: this gives raise to the
presence of off-diagonal terms in the metric tensor, which are
responsible for a variety of effects concerning orbiting test
particles, precessing gyroscopes, moving clocks and atoms and
propagating electromagnetic waves
\citep{mashh1,ruggiero02,Scia,Mash}. They are expected in GTR, but
are generally very small and, hence, very difficult to detect
\citep{Ior07}. In recent years, there have been some attempts to
measure the Lense-Thirring effect \citep{LT} with the LAGEOS and
LAGEOS II laser-ranged satellites in the gravitational field of
the Earth \citep{ciufolini04}; the evaluation of the realistic
accuracy reached in such a test and other topics related to it are
still matter of debate \citep{IorJoG,IorPss}. For other attempts
to measure the Lense-Thirring effect in other Solar System
scenarios with natural and artificial satellites, see
\citep{Ior07}. In April 2004 the Gravity Probe B spacecraft
\citep{gpb} was launched to accurately measure the
gravito-magnetic (and  geodetic) precession of an orbiting
gyroscope \citep{Pug,Sch} in the terrestrial space environment:
the final results are going to be published. We focus on the GM
effects in Kerr-de Sitter metric (the GM precession of an orbiting gyroscope was
investigated by \citet{ruggiero08}): in particular we  work out
the GM effects in the weak-field and slow-motion approximation on
the orbit of a test particle, working out explicitly the
perturbations of the Keplerian orbital elements; furthermore, by
using the EPM2004 \citep{Pit05a} and EPM2006 \citep{EPM2006}
ephemerides, we put constraints on the parameter $k$, which is the cosmological constant in  GTR and a
measure of the non linearity of the theory in $f(R)$ gravity.

\section{Gravito-magnetic field in Kerr-de Sitter metric}
\label{sec:gmfr}

The Kerr-de Sitter metric in the standard Boyer-Lindquist
coordinates $x^{\mu} = (t, \rho, \theta, \phi)$ has the form\footnote{The space-time metric has signature $(1,-1,-1,-1)$,
greek indices run from 0 to 3, and latin ones run from 1 to 3,
boldface letters like $\bm{r}$ refers to three-vectors.}

\begin{eqnarray} ds^2=& & \left\{ 1 - \frac{2GM\rho}{c^2\Sigma} - \frac{k }{3} \left[\rho^2 + \left(\frac{J}{Mc}\right)^2\; {\rm
sin}^2\theta\right]\right\} c^2dt^2  + \nonumber\cr\\
& + & 2\left(\frac{J}{Mc}\right)\left\{\frac{2GM\rho}{c^2\Sigma} +
\frac{k }{3} \left[\rho^2 + \left(\frac{J}{Mc}\right)^2\right]\right\}\;{\rm
sin}^2\theta c dt d\phi + \frac{\Sigma}{\Delta} d\rho^2 +
\frac{\Sigma}{\chi} d\theta^2 + \nonumber\cr\\ &+& \left\{
\frac{2GM\rho}{c^2\Sigma} \left(\frac{J}{Mc}\right)^2{\rm
sin}^2\theta + \left[1+ \frac{k }{3} \left(\frac{J}{Mc}\right)^2\right]
\left[\rho^2 + \left(\frac{J}{Mc}\right)^2\right]\right\} \; {\rm sin}^2\theta
d\phi^2\;\;, \label{eq:kdsmetric1}
\end{eqnarray}
where
\begin{equation}
\Sigma = \rho^2+
\left(\frac{J}{Mc}\right)^2\cos^2\theta\;\;,\;\;\chi = 1 + \frac{k
}{3} \left(\frac{J}{Mc}\right)^2\cos^2\theta\;\;,
\label{eq:kdsdef1}
\end{equation}
\begin{equation}
\Delta = \rho^2 - 2\frac{GM\rho}{c^2} +
\left(\frac{J}{Mc}\right)^2 - \frac{k }{3} \rho^2\left[\rho^2 +
\left(\frac{J}{Mc}\right)^2\right]\;\;. \label{eq:kdsdef2}
\end{equation}
The mass of the source is $M$, while $J$ is its angular momentum
(which is perpendicular to the $\theta=\pi/2$ plane); $k$ is the cosmological constant in  GTR framework,
while in $f(R)$ theories \citep{ruggiero08} it is a parameter related to the non linearity of the gravity lagrangian (namely, when $f(R)=R$ then $k=0$).
When $k=0$ the Kerr-de Sitter metric given by \rfr{eq:kdsmetric1} reduces to the
Kerr metric. Other limiting cases can be checked: for instance,
when $J=0$, we obtain the Schwarzschild-de Sitter solution, and
when $M=J=0$ we have the de Sitter space-time.

In order to study GM effects a weak field approximation \rfr{eq:kdsmetric1} is sufficient; furthermore, it is
useful to introduce the  isotropic radial coordinate $r$ denotes
the defined as \eqI r = \rho\left(1-\rp{GM}{c^2\rho} -
\rp{k\rho^2}{12}\right),\eqF where $\rho$ is the standard
Boyer-Lindquist radial coordinate. Then, up to linear terms in
$\rp{GM}{c^2 r}$, $\rp{GJ}{c^3 r^{2}}$, $kr^{2}$, $\rp{kJr}{cM}$, the metric is

\begin{align}
ds^2=&\left(1-\frac{2GM}{c^2r}-\frac{k }{3}r^2\right)c^2dt^2 -
\left(1+\frac{2GM}{c^2r} -\frac{k }{6}r^2
\right)\left(dr^2+r^2 d\theta^2+r^2 \sin^2 \theta d\phi^2\right)+ \notag \\
& +2\frac{J}{Mc}\left[\frac{2GM}{c^2r}+\frac{kr }{3}\left(r+\frac{5}{2}
\frac{GM}{c^2} \right)  \right]\sin^2  \theta d\phi c dt.
\label{eq:kdsweak2}
\end{align}

By differentiating
\eqI \rp{y}{x} = \tan\phi \eqF and using
\eqI \cos\phi = \rp{x}{r\sin\theta},\eqF it  turns out that
\eqI \sin^2\theta d\phi = \rp{-y\ dx + x\ dy}{r^2};\eqF
thus,
the off-diagonal, i.e.
gravito-magnetic, components $g_{0i},\ i=1,2,3$ of the metric
tensor of the weak-field approximation of the Kerr-de Sitter
space-time are, in cartesian coordinates,
\begin{eqnarray}
  g_{01} &=& -\rp{J}{Mc}\left[\rp{2GM}{c^2 r^3} + \rp{k}{3}\left(1 + \rp{5GM}{2c^2 r}\right)\right]{y}, \\
  g_{02} &=&  \rp{J}{Mc}\left[\rp{2GM}{c^2 r^3} + \rp{k}{3}\left(1 + \rp{5GM}{2c^2 r}\right)\right]{x} \\
  g_{03} &=& 0;
\end{eqnarray}

In the weak-field and slow-motion linear approximation
the spatial components of the geodesic equations of motions yielding gravito-magnetic accelerations are \citep{Bru}
\eqI\rp{d^2x^i}{dt^2}= c\left(\partial_j h_{0i}-\partial_i h_{0j}\right)\dert {x^j} t,\ i=1,2,3\eqF
where $h_{0l}=g_{0l}-\eta_{0l}=g_{0l},\ l=1,2,3$. It can be straightforwardly  showed that the terms not containing $k$ yield the usual Lense-Thirring acceleration in cartesian coordinates \citep{Sof}.
The components of the acceleration containing $k$ in cartesian coordinates are
\begin{eqnarray}\lb{acc_cart}
  A_x &=& \rp{2GJk}{3c^2}\left\{  -\left[\rp{c^2}{GM} + \rp{5}{4}\left(\rp{x^2 + y^2+ 2z^2}{r^3}\right)  \right]\dot y +\left(\rp{5yz}{4r^3}\right)\dot z   \right\}, \\
  A_y &=& \rp{2GJk}{3c^2}\left\{  \left[\rp{c^2}{GM} + \rp{5}{4}\left(\rp{x^2 + y^2+ 2z^2}{r^3}\right)  \right]\dot x -\left(\rp{5xz}{4r^3}\right)\dot z   \right\},\\
  A_z &=& \rp{5GJk}{6c^2}\left[\rp{z(x\dot y - y\dot x)}{r^3}\right].
\end{eqnarray}
which can be cast into the vectorial form  \citep{Mash} \eqI \bds
A= -2\rp{{\bds v}}{c}\bds\times {\bds{\mathcal{B}}},\lb{acc}\eqF where
the gravito-magnetic field $\bds{\mathcal{B}}$ is  \eqI
\bds{\mathcal{B}} = \rp{Jkc}{3M}\bds{\hat{J}} +
\rp{5GJk}{12c}\rp{\left[\bds{\hat{J}} +
\left(\bds{\hat{J}}\bds\cdot\bds{\hat{r}}\right)\bds{\hat{r}}\right]}{r},\eqF
with $\bds{\hat{J}}=\bds{\hat{z}}$; the unit vector in the radial direction is defined as
\eqI \bds{\hat{r}}=\rp{\bds r}{r}.\eqF We notice that the
gravito-magnetic field consists of two contributions, the first
one that is everywhere constant and parallel to $\bds J$, the
second one whose position and directions are position-dependent.

Furthermore, by defining
the gravito-magnetic potential $\bds{\mathcal{A}}$ as \citep{Mash} \eqI
\bds{\mathcal{A}} = kr^2\left(\rp{c^2 r}{6GM}+\rp{5}{12}\right)\rp{G}{c}\rp{\bds
{J}\bds\times \bds r}{r^3}\eqF and using
\eqI \bds\nabla\bds\times(\bds {\rm A}\bds\times \bds {\rm B})=(\bds {\rm B}\bds\cdot\bds\nabla)\bds {\rm A}- (\bds {\rm A}\bds\cdot\bds\nabla)\bds {\rm B} + \bds {\rm A}(\bds\nabla\bds\cdot \bds {\rm B}) - \bds {\rm B}(\bds\nabla\bds\cdot \bds {\rm A}),\eqF
\eqI \bds{\rm A}\bds\times\left(\bds{\rm B}\bds\times\bds{\rm C}\right) = \left(\bds{\rm A}\bds\cdot\bds{\rm C}\right)\bds{\rm B} - \left(\bds{\rm A }\bds\cdot\bds{\rm B}\right)\bds{\rm C},\eqF
\eqI\bds\nabla\bds\times\left(\psi \bds {\rm H}\right)=\bds\nabla\psi\bds\times\bds {\rm H}+\psi\bds\nabla\bds\times\bds {\rm H},\eqF
\eqI\bds\nabla\bds\cdot \bds r = 3, \eqF
\eqI \bds\nabla\rp{1}{r}=-\rp{\bds{\hat{r}}}{r^2}\eqF
with $\psi = 1/r,$
it is possible to express the gravito-magnetic field $\bds{\mathcal{B}}$ in terms of $\bds{\mathcal{A}}$ as \citep{Mash}\eqI
\bds{\mathcal{B}}=\bds\nabla\bds\times \bds{\mathcal{A}}.\eqF

In order to calculate the impact of \rfr{acc} on the orbit of a test particle, let us project it onto the radial ($\bds{\hat{r}}$), transverse ($\bds{\hat{t}}$) and normal ($\bds{\hat{n}}$) directions of the co-moving frame picked out by the three unit vectors\footnote{Here we have chosen the $x$ axis coincident with the line of the nodes, i.e. $\Omega=0$.}  \citep{Mont}
\begin{eqnarray}\lb{tripod}
  \bds{\hat{r}} &=& \cos u\ \bds{\hat{x}} + \cos i\sin u\ \bds{\hat{y}} + \sin i\sin u\ \bds{\hat{z}}, \\
  \bds{\hat{t}} &=& -\sin u\ \bds{\hat{x}} + \cos i\cos u\ \bds{\hat{y}} + \sin i\cos u\ \bds{\hat{z}} \\
  \bds{\hat{n}} &=& -\sin i\ \bds{\hat{y}} + \cos i\ \bds{\hat{z}},
\end{eqnarray}
where $i$ is the inclination of the orbital plane to the equator of the central mass and $u=\omega+f$ is the argument of latitude defined as the sum of the argument of the pericentre $\omega$, which fixes the position of the pericentre with respect to the line of the nodes, and the true anomaly $f$ which reckons the position of the test particle from the pericentre.
Thus,
\begin{eqnarray}
  A_r &=& \bds A\bds\cdot \bds{\hat{r}}\ =\ -\rp{GJk}{6c^2}\left(5 + 4\rp{c^2 r}{GM}\right)\cos i\ \dot u, \lb{Acc_rtn} \\
  A_t &=& \bds A\bds\cdot \bds{\hat{t}}\ =\ 0, \\
  A_n &=& \bds A\bds\cdot \bds{\hat{n}}\ =\ \rp{GJk}{3 c^2}\left(5 + 2\rp{c^2 r}{GM}\right)\sin i\sin u\ \dot u\lb{Acc_rtn2};
\end{eqnarray}
they must be inserted  into the right-hand-side of the Gauss equations \citep{Ber} of the variations of the Keplerian orbital elements
\begin{eqnarray}\lb{Gauss}
\dert a t & = & \rp{2}{n\sqrt{1-e^2}} \left[e A_r\sin f +A_{t}\left(\rp{p}{r}\right)\right],\lb{gaus_a}\\
\dert e t  & = & \rp{\sqrt{1-e^2}}{na}\left\{A_r\sin f + A_{t}\left[\cos f + \rp{1}{e}\left(1 - \rp{r}{a}\right)\right]\right\},\lb{gaus_e}\\
\dert i t & = & \rp{1}{na\sqrt{1-e^2}}A_n\left(\rp{r}{a}\right)\cos u,\\
\dert\Omega t & = & \rp{1}{na\sin i\sqrt{1-e^2}}A_n\left(\rp{r}{a}\right)\sin u,\lb{gaus_O}\\
\dert\omega t & = &\rp{\sqrt{1-e^2}}{nae}\left[-A_r\cos f + A_{t}\left(1+\rp{r}{p}\right)\sin f\right]-\cos i\dert\Omega t,\lb{gaus_o}\\
\dert {\mathcal{M}} t & = & n - \rp{2}{na} A_r\left(\rp{r}{a}\right) -\sqrt{1-e^2}\left(\dert\omega t + \cos i \dert\Omega t\right),\lb{gaus_M}
\end{eqnarray}
where $a$, $e$, $\Omega$ and ${\mathcal{M}}$ are the semi-major axis, the eccentricity, the longitude of the ascending node and the mean anomaly of the orbit of the test particle, respectively,  $p=a(1-e^2)$ is the semi-latus rectum and $n=\sqrt{GM/a^3}$ is the un-perturbed Keplerian mean motion.
By evaluating them onto the un-perturbed Keplerian ellipse
\eqI r = \rp{a(1-e^2)}{1+e\cos f}\eqF    and averaging\footnote{We used $\dot u=\dot f$ because over one orbital revolution the pericentre $\omega$ can be assumed constant.} them over one orbital period $P_{\rm b}$ of the test particle  by means of
\eqI \rp{dt}{P_{\rm b}} = \rp{(1-e^2)^{3/2}}{2\pi(1+e\cos f)^2}df,\eqF
it is possible to obtain the secular effects induced by \rfr{acc}
\begin{eqnarray}\lb{orbi}
  \left\langle\dot a\right\rangle &=& 0, \\
  \left\langle\dot e\right\rangle &=& 0, \\
  \left\langle\dot i\right\rangle &=& 0, \\
  \left\langle\dot\Omega\right\rangle &=& \rp{Jk}{3M}\left(1+\rp{5GM}{2c^2 a}\right), \lb{k_O}\\
  \left\langle\dot\omega\right\rangle &=& -\rp{Jk\cos i}{3M}\left(2 +\rp{5GM}{2c^2 a}\right),\lb{k_o} \\
  \left\langle\dot{\mathcal{M}}\right\rangle &=& n + \rp{5Jk\cos i}{3M}\left(1 + \rp{GM}{c^2 a}\right).
\end{eqnarray}
In the calculation we have neglected terms of order $\mathcal{O}(e^2)$.

The correction $\Delta P_{\rm b}$ to the orbital period $P_{\rm b}$ due to \rfr{acc} can be calculated using the mean longitude
\eqI\lambda = \mathcal{M}+\omega +\cos i\ \Omega.\eqF
For small eccentricities \rfr{gaus_O}-\rfr{gaus_M} yield
\eqI \dert\lambda t \approx n -\rp{2}{na} A_r\left(\rp{r}{a}\right).\eqF By using \rfr{Acc_rtn} it is possible to obtain, for $e\rightarrow 0$,
\eqI P_{\rm b} \approx \rp{2\pi}{n}\left[1-\rp{GJk}{3c^2 n a}\left(\rp{4c^2 a}{GM} +5\right)\cos i\right],\eqF so that
\eqI \Delta P_{\rm b} = -\rp{2\pi Jka^2}{3c^2 M}\left(\rp{4c^2 a}{GM}+5\right)\cos i.\eqF

We will, now, put constraints on $k$ from the corrections to the standard Newtonian/Einsteinian precessions of the longitudes of the perihelia $\varpi=\omega+\cos i\ \Omega$
of the inner planets of the Solar System, quoted in Table \ref{tavola}, estimated by E.V. Pitjeva by fitting more than 400000 observations of various kinds with the EPM2004 \citep{Pit05a,Pit05b} and EPM2006 \citep{EPM2006} ephemerides. No gravito-magnetic terms of any kind were included in the dynamical force models used, so that, in principle, they account for the effects investigated by us.
\small\begin{table*}
\small
\caption{ Inner planets. First row: estimated perihelion
extra-precessions in $10^{-4}$ \asec\ (\asec$\rightarrow$ arcseconds per century), from Table 3 of \citep{Pit05b} (apart from Venus). The quoted
errors, in $10^{-4}$ \asec, are not the formal ones but are  realistic. The formal errors  are quoted in square brackets (E.V. Pitjeva, personal communication to L.I., November 2005). The units are
$10^{-4}$ \asec. Second row: semi-major axes, in Astronomical Units (AU). Their formal errors are in Table IV of \citep{Pit05a}, in m. Third row: eccentricities. Fourth row: orbital periods in years. The result for Venus have been recently obtained by including the Magellan radiometric data (E.V. Pitjeva, personal communication to L.I., June 2008).\label{tavola} }

\small\begin{tabular}{@{}lllll@{}}
\hline
& Mercury & Venus & Earth & Mars\\
\tableline
$\left<\Delta\dot\varpi\right>$ ($10^{-4}$\ \asec) & $-36\pm 50[42]$ & $-4\pm 5[1]$& $-2\pm 4[1]$ & $1\pm 5[1]$\\
$a$ (AU) & 0.387 & 0.723 & 1.000 & 1.523 \\
$e$ & 0.2056 & 0.0067 & 0.0167  & 0.0934\\
$P$ (yr) & 0.24 & 0.61 &  1.00 & 1.88\\
\hline

\end{tabular}

\end{table*}
\normalsize For the spin angular momentum of the Sun we will use
the value $J_{\odot}=(190.0\pm 1.5)\times 10^{39}$ kg m$^2$
s$^{-1}$ determined from helioseismology \citep{Pij98,Pij03}, i.e.
independently of the planetary dynamics which we want to test.
From \eqI \left\langle\dot\varpi\right\rangle=-\rp{Jk\cos
i}{3M},\eqF and Table \ref{tavola} it is possible to obtain \eqI
k\leq 1\times 10^{-29}\ {\rm m}^{-2}.\eqF

In the case of the laser-ranged LAGEOS satellite \citep{LAGE}, orbiting at about 6000 km above the Earth's surface, by assuming $J_{\oplus}=5.85\times 10^{33}$ kg m$^2$ s$^{-1}$ \citep{IERS} and an uncertainty of the order of 1 cm or less \citep{prec} in reconstructing its orbit, which translates  into an uncertainty in the nodal rate of $\delta\dot\Omega\sim 0.1$ milliarcseconds per year, the bound which can be obtained from \rfr{k_O} is $k\leq 4\times 10^{-26}$ m$^{-2}$.

Concerning the double pulsar system PSR J0737-3039A/B \citep{Bur}, by assuming for the moment of inertia of A the value
$I\approx 10^{38}$ kg m$^2$ \citep{lor}, since its rotational period is 22 ms \citep{Kra06} its angular momentum can be evaluated as $J_{\rm A}=2.8\times 10^{40}$ kg m$^2$ s$^{-1}$. The overall uncertainty (including also the mismodelling in the usual 1PN term) in the periastron precession  amounts to 0.03 deg yr$^{-1}$ \citep{Ior08}, so that \rfr{k_o} and the system's parameters \citep{Kra06} yield $k\leq 3\times 10^{-21}$ m$^{-2}$.

In order to evaluate such bounds it is useful remember that studying the Schwarzschild-de Sitter non-gravitomagnetic precession various authors \citep{kagramanova06,IorioL,Sereno,Jetzer,Ruggiero} have obtained bounds on the cosmological constant that range from $10^{-40}$ m$^{-2}$ to about $10^{-42}$ m$^{-2}$; furthermore, the data in the general relativistic $\Lambda$CDM (cosmological constant plus cold dark matter) model suggest that a value of the cosmological constant of $10 ^{-52}$ m$^{-2}$ is required in order to explain the accelerated expansion of the Universe \citep{cosmocost}.   As a consequence, the bounds that we have just obtained are not competitive. The same conclusions apply to Palatini $f(R)$ gravity: in fact when the formation of large-scale structure is studied in this context (i.e. when the non linear part of $f(R)$ is supposed to drive cosmic acceleration) the results are practically indistinguishable from the $\Lambda$CDM model; see e.g. \citet{koivisto1} who used large scale structure cosmological data from Sloan Digital Sky Survey (SDSS). However, these tight constraints on $f(R)$ gravity are loosened when dark matter  with inherent stresses  (generalized dark matter, GDM) is allowed \citep{koivisto}.

\section{Discussion and Conclusions} \label{sec:disconc}
We explicitly worked out the effects induced on the orbit of a
test particle by the weak-field approximation of the Kerr-de
Sitter metric, which is a solution of the vacuum field equations both in  GTR and in $f(R)$ gravity.
It turns out that the semi-major axis,
the eccentricity and the inclination do not experience secular,
i.e. averaged over one orbital period, changes; instead, the
longitude of the ascending node, the argument of pericentre and
the mean anomaly undergo secular precessions.
Interestingly, all such
effects consist of two kinds of contributions. The first type is
given by terms proportional to $GMc^{-2}$, which vanish in the
limits $c\rightarrow\infty$, $G\rightarrow 0$, $M\rightarrow 0$.
Instead, the second kind consists of terms proportional to $Jk/M$,
which are independent of the speed of light $c$, the constant of
gravitation $G$ and the source's mass $M$, so that they do not
vanish in the limits for $c\rightarrow\infty$, $G\rightarrow 0$
and  $M\rightarrow 0$. Concerning the dependence on the orbital
geometry, both kinds of effects depend on the inclination and
vanish for polar orbits; while the $\mathcal{O}(c^{-2})$ terms
depend also on the size of the orbit through the semi-major axis,
it is not so for the $\mathcal{O}(Jk/M)$ ones which are, indeed,
independent of it.
Then, we compared our predictions to the latest observational
determinations of the corrections to the standard
Newtonian/Einsteinian precessions of the perihelia of the inner
planets of the Solar System obtaining the constrain $k\leq
10^{-29}$ m$^{-2}$. The node of the terrestrial LAGEOS satellite  yields $k\leq
10^{-26}$ m$^{-2}$, while the bound from the periastron of the double pulsar system PSR J0737-3039A/B is
$k\leq 3\times 10^{-21}$ m$^{-2}$.
Such bounds are not competitive with the ones which can be obtained from the Schwarzschild-de Sitter non-gravitomagnetic precessions ($k\leq 9\times 10^{-43}$ m$^{-2}$ from Solar System data) and from those deriving from cosmological observations.

\end{document}